\documentclass[journal,twoside,final]{IEEEtran}

\usepackage[boxed]{algorithm2e}
\usepackage{amsmath}
\usepackage{amssymb}
\usepackage{balance}
\usepackage{cite}
\usepackage{graphicx}
\usepackage{multirow}
\usepackage{threeparttable}
\usepackage{verbatim}

\graphicspath{{./}}

\begin{document}
\title{Atomic Scheduling of Appliance Energy Consumption in Residential Smart
  Grid}

\author{%
  Kyeong Soo Kim, \IEEEmembership{Member, IEEE}, Sanghyuk Lee, Tiew On Ting,
  \IEEEmembership{Member, IEEE}, and Xin-She Yang%
  \thanks{%
    K. S. Kim, S. Lee, and T. O. Ting are with the Department of Electrical and
    Electronic Engineering, Xi'an Jiaotong-Liverpool University, Suzhou 215123,
    Jiangsu Province, P. R. China (e-mail:
    \{Kyeongsoo.Kim,Sanghyuk.Lee,Toting\}@xjtlu.edu.cn).%
  }
  \thanks{%
    Xin-She Yang is with the School of Science and Technology, Middlesex
    University, London NW4 4BT, United Kingdom (e-mail: x.yang@mdx.ac.uk).%
  }%
}


\maketitle

\begin{abstract}
  The current formulation of the optimal scheduling of appliance energy
  consumption uses as optimization variables the vectors of appliances'
  scheduled energy consumption over equally-divided time slots of a day, which
  does not take into account the \textit{atomicity} of appliances' operations
  (i.e., the unsplittable nature of appliances' operations and resulting energy
  consumption). In this paper, we provide a new formulation of atomic scheduling
  of energy consumption based on the \textit{optimal routing} framework; the
  flow configurations of users over multiple paths between the common source and
  destination nodes of a ring network are used as optimization variables, which
  indicate the starting times of scheduled energy consumption, and optimal
  scheduling problems are now formulated in terms of the user flow
  configurations. Because the atomic optimal scheduling results in a
  \textit{Boolean-convex} problem for a convex objective function, we propose a
  \textit{successive convex relaxation} technique for efficient calculation of
  an approximate solution, where we iteratively drop fractional-valued elements
  and apply convex relaxation to the resulting problem until we find a feasible
  suboptimal solution. Numerical results for the cost and peak-to-average ratio
  minimization problems demonstrate that the successive convex relaxation
  technique can provide solutions close to, often identical to, global optimal
  solutions.
\end{abstract}


%

\section{Introduction}
\label{sec:introduction}
\IEEEPARstart{W}{e} study the problem of scheduling electrical appliance energy
consumption in residential smart grid. Our goal in revisiting this well-known
problem of energy consumption scheduling (e.g., \cite{mohsenian-rad10:_auton,
  logenthiran12:_deman_side_manag_smart_grid, adika12:_auton,
  baharlouei13:_achiev_optim_fairn_auton_deman_respon, hu13:_multiob,
  baharlouei14:_effic_fairn_trade_privac_preser,
  liu14:_peak_averag_ratio_const_deman,
  khan14:_peak_load_sched_smart_grid_commun_envir, park15:_bargain}) is to
establish a new formulation of optimal scheduling problems where we take into
account the \textit{atomicity} of operations by household appliances, and to
provide efficient solution techniques for the formulated optimal scheduling
problems. By atomicity, we mean the \textit{unsplittable} nature of appliances'
operations and resulting energy consumption.

Note that the scheduling of electrical appliance energy consumption is a key to
the autonomous demand-side management (DSM) for residential smart grid in
optimizing energy production and consumption; the scheduling is based on smart
meters installed at users' premises and the two-way digital communications
between a utility company and users through the smart meters, and the typical
goals of DSM includes consumption reducing and shifting, which lead into lower
peak-to average ratio (PAR) and energy cost \cite{saad12:_game}.

Since the energy consumption scheduling was formulated as an optimization
problem using energy consumption scheduling vectors as optimization variables
representing appliances' scheduled hourly energy consumption over a day
\cite{mohsenian-rad10:_auton}, there have been published a number of papers on
the subjects of appliance energy consumption scheduling and related
billing/pricing mechanisms based on this formulation. For instance, the issue of
optimality and fairness in autonomous DSM is studied in relation with billing
mechanisms in \cite{baharlouei13:_achiev_optim_fairn_auton_deman_respon}, while
the same issue is studied but in the context of user privacy in
\cite{baharlouei14:_effic_fairn_trade_privac_preser}. The cost and PAR
minimization problems, which are separately formulated in
\cite{mohsenian-rad10:_auton}, are integrated into a PAR-constrained cost
minimization problem in \cite{liu14:_peak_averag_ratio_const_deman}. This
integration is also further extended to take into account consumers' preference
on operation delay and power gap using multiple objective functions. In
\cite{park15:_bargain}, instead of typical concave $n$-person games,
Rubinstein-Stahl bargaining game model is used to capture the interaction
between the supplier and the consumers through a retail price vector in lowering
PAR to a certain desired value.

In most existing works on the energy consumption scheduling for autonomous DSM
in smart grid, however, no serious attention has been given to the
microstructure of scheduled energy consumption over time slots; their major
focus is on the optimal value of an objective function that is only based on the
aggregate load from the scheduled energy consumption. Few exceptions in this
regard include the works on the integration of consumers' preference
\cite{liu14:_peak_averag_ratio_const_deman} and the use of load consumption
curves in the objective function
\cite{logenthiran12:_deman_side_manag_smart_grid}.

Below are some scenarios illustrating the importance of the atomicity of
appliance operations.
\begin{itemize}
\item When a user washes clothes, the washing machine should be continuously on
  for a certain period depending on the amount of laundry, e.g., for two hours,
  during which the operation of the clothes washer cannot be interrupted and the
  supplied power cannot be reduced arbitrarily. As the washing task can be
  activated anytime within a specified period, e.g., from 9 AM to 3 PM when the
  user is out to work, the major goal of autonomous DSM is to determine the
  optimal two-hour time slot to complete the task when the energy price is
  lowest during the specified period.
\item For heavier-duty tasks like charging the battery of a plug-in hybrid
  electric vehicle (PHEV), again the continuity of the operation is important,
  e.g., four hours of uninterruptible charging to maintain the lifetime of
  PHEV's battery.
\item Appliances such as a rice cooker can significantly contribute to the
  overall cost saving if handled properly. Some rice cookers' function may take
  more than one hour for completion, e.g., the slow cooking function for
  delicate soup, whereby the cooking process can be scheduled within any
  time-shot during the day.
\end{itemize}

Considering that most of the operations subject to DSM are either atomic as such
(e.g., laundry cleaning by a clothes washer) or consist of atomic suboperations
(e.g., house heating by a heater operating in the morning and in the evening)
\cite{khan14:_peak_load_sched_smart_grid_commun_envir}, therefore, we provide a
new formulation of the optimization problem for atomic scheduling of appliance
energy consumption and efficient solution techniques for resulting problems in
this paper. Atomic scheduling has been mostly discussed in the context of
concurrent task scheduling on a multi-processor/core system on a chip (SOC)
(e.g., \cite{yang01:_energ_socs}) or transaction processing (e.g.,
\cite{stankovic88:_real_trans}). To the best of our knowledge, our work is the
first attempt to formulate atomic scheduling of appliance energy consumption in
the autonomous DSM for residential smart grid.

The rest of the paper is organized as follows: In
Section~\ref{sec:curr-form-appl}, we review and discuss the issues of the
current formulation of appliance energy consumption scheduling based on energy
consumption scheduling vectors defined over time slots. In
Section~\ref{sec:atom-sched-appl}, we describe a new formulation of the
appliance energy consumption scheduling based on the optimal routing framework,
which guarantees the atomicity of appliance operations, and a successive convex
relaxation technique for the efficient solution of the Boolean-convex problem
resulting from the new formulation for a given convex objective function. In
Section~\ref{sec:numerical_results}, we demonstrate that the performance of
successive convex relaxation technique through numerical results for the cost
and PAR minimization problems. Section~\ref{sec:conclusions} concludes our work
and discusses topics for further study.

\section{Review of Current Formulation of Appliance Energy Consumption
  Scheduling}
\label{sec:curr-form-appl}
We first review the current formulation of appliance energy consumption
scheduling problem by formally describing it. The formulation described here is
largely based on \cite{mohsenian-rad10:_auton} but with some modifications and
extensions for clarity and better handling of scheduling intervals over a day
boundary. Many notations and definitions in this section are applicable to the
formulation of atomic scheduling problems in Section~\ref{sec:atom-sched-appl}
as well.

Let $\mathcal{N}{\triangleq}\left\{1,{\ldots},N\right\}$ denotes a set of users
in a residential smart grid, whose appliances share a common energy source and
subject to autonomous DSM. Without loss of generality and for ease of
presentation, we assume that each user has only one appliance throughout the
paper.\footnote{We use the term ``user'' and ``appliance'' interchangeably.} In
this case a daily energy consumption scheduling vector of user $n$ is defined as
\begin{equation}
  \label{eq:energy_consumption_vector}
  \mathbf{x}_{n} \triangleq \left[x^0_{n},\ldots,x^h_{n},\ldots,x^{H-1}_{n}\right]
\end{equation}
where a scalar element $x^h_{n}$ denotes the energy consumption scheduled for a
time slot $h{\in}\mathcal{H}{\triangleq}\left\{0,{\ldots},H{-}1\right\}$.
\footnote{Time slot numbering in this paper starts from 0, which makes it easier
  handling energy consumption scheduling wrap-around the day boundary (i.e.,
  from 11 PM to 6 AM) using modulo operation. See
  \eqref{eq:non-atomic_feasible_set} and
  \eqref{eq:non-atomic_scheduling_interval} for details.}
A feasible energy consumption scheduling set for user $n$ is given by
\begin{equation}
  \label{eq:non-atomic_feasible_set}
  \begin{split}
    \mathcal{X}_n{=} & \Bigg\{\mathbf{x}_n \Bigg|
    \sum_{h \in \mathcal{H}_{n}}x^h_{n}{=}E_{n},\\
    & \gamma^{min}_{n}{\leq}x^h_{n}{\leq}\gamma^{max}_{n},~\forall h {\in}
    \mathcal{H}_{n},~x^h_{n}{=}0,~\forall h {\in} \mathcal{H} \backslash
    \mathcal{H}_{n}\Bigg\}
  \end{split}
\end{equation}
where $\gamma^{min}_{n}$ and $\gamma^{max}_{n}$ are the minimum and the maximum
energy levels for a time slot, $E_{n}$ is the total daily energy consumption of
user $n$'s appliance, and $\mathcal{H}_{n}$ is a \textit{scheduling interval}
defined as follows
\begin{equation}
  \label{eq:non-atomic_scheduling_interval}
  \mathcal{H}_{n} \triangleq \left\{h \big|h=i \bmod H,~\forall i{\in} \left[\alpha_{n},~\beta_{n}\right]\right\}
\end{equation}
with $\alpha_{n}{\in}[0,~H{-}1]$, $\beta_{n}{\in}[1,~2H{-}2]$, and
$1{\leq}\beta_{n}{-}\alpha_{n}{\leq}H{-}1$.

With these definitions of scheduling vectors and feasible sets, the optimal
scheduling is formulated as an optimization problem for a given objective
function (e.g., total energy cost or PAR) as follows:
\begin{equation}
  \label{eq:optimal_scheduling}
  \underset{\mathbf{x}_n \in \mathcal{X}_{n},~\forall n \in
    \mathcal{N}}{\mathbf{minimize}} ~ \psi(\mathbf{x})
\end{equation}
where
\begin{equation}
  \mathbf{x} \triangleq \left[\mathbf{x}_{1},\ldots,\mathbf{x}_{N}\right] .
\end{equation}

As shown in \cite{mohsenian-rad10:_auton}, it is better to develop a DSM
approach that optimizes the properties of the aggregate load of the users rather
than individual user's consumption. In fact, the optimal scheduling problems are
formulated in terms of the total load across all users at each time slot $h$ in
\cite{mohsenian-rad10:_auton}, i.e.,
\begin{equation}
  L_{h}\left(\mathbf{x}\right) \triangleq \sum_{n \in \mathcal{N}} x_{n}^{h} .
\end{equation}
So \eqref{eq:optimal_scheduling} can be expressed as
\begin{equation}
  \label{eq:optimal_scheduling_aggregated}
  \underset{\mathbf{x}_n \in \mathcal{X}_{n},~\forall n \in
    \mathcal{N}}{\mathbf{minimize}} ~ \phi\left(\boldsymbol{L}(\mathbf{x})\right)
\end{equation}
where
\begin{equation}
  \boldsymbol{L}(\mathbf{x}) \triangleq
  \left[L_{0}\left(\mathbf{x}\right),\ldots,L_{H-1}\left(\mathbf{x}\right)\right].
\end{equation}
Note that the objective function becomes
\begin{equation}
  \phi\left(\boldsymbol{L}(\mathbf{x})\right) = \sum_{h \in \mathcal{H}} C_{h}\left(L_{h}(\mathbf{x})\right)
\end{equation}
for the energy cost minimization problem, where $C_{h}(\cdot)$ is a cost
function indicating the cost of generating or distributing electricity energy by
the energy source at a time slot $h$, and
\begin{equation}
  \phi\left(\boldsymbol{L}(\mathbf{x})\right) = \frac{\displaystyle H \max_{h \in \mathcal{H}}
    L_{h}(\mathbf{x})}{\displaystyle \sum_{n \in \mathcal{N}} E_{n}}
\end{equation}
for the PAR minimization problem, respectively.

\begin{figure}[!tb]
  \begin{center}
    \includegraphics[width=.8\linewidth]{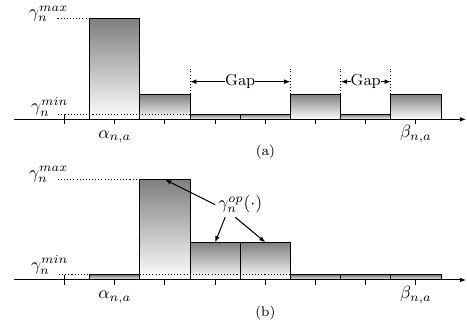}%
  \end{center}
  \addtolength\abovecaptionskip{-10pt} 
  \caption{Examples of (a) non-atomic and (b) atomic scheduling of appliance
    energy consumption, where $\gamma^{min}_{n}$, $\gamma^{max}_{n}$, and
    $\gamma^{op}_{n}(\cdot)$ are the minimum, the maximum, and the operating
    energy levels of an appliance for a time slot.}
  \label{fig:non-atomic_vs_atomic}
\end{figure}

A major issue with the formulation of optimal scheduling problems based on the
energy consumption scheduling vectors in \eqref{eq:energy_consumption_vector} is
that it cannot guarantee the atomicity of appliance operations, which is
illustrated in Fig.~\ref{fig:non-atomic_vs_atomic}~(a). When a load from other
appliances falls on the middle of the scheduling interval, especially that of
non-shiftable appliances, the scheduled appliance energy consumption spreads
over non-contiguous time slots, resulting in several gaps. Relatedly, the
scheduled energy consumption may not provide enough power for appliances to
carry out required operations because, during the scheduling, the energy levels
over time slots are determined to achieve the optimal solution for the given
objective function but not to meet the actual energy consumption requirements of
the appliances for the operations. The atomic scheduling that will be described
in Section~\ref{sec:atom-sched-appl}, on the other hand, assigns contiguous time
slots with a predefined pattern of operating energy levels (i.e.,
$\gamma^{op}_{n}(\cdot)$) as shown in Fig.~\ref{fig:non-atomic_vs_atomic}~(b),
even at the expense of increased penalty in optimization.

\section{Atomic Scheduling of Appliance Energy Consumption}
\label{sec:atom-sched-appl}
\begin{figure}[!tb]
  \begin{center}
    \includegraphics[width=.8\linewidth]{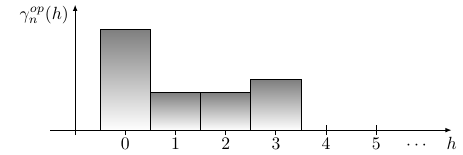}%
  \end{center}
  \addtolength\abovecaptionskip{-10pt} 
  \caption{An example of a predefined pattern of operating energy levels for
    $\delta_{n}=4$.}
  \label{fig:op_energy_level_example}
\end{figure}

\noindent
We assume that all appliance operations are atomic (i.e., no gaps in energy
consumption during the operations) and that the operation of the appliance of
user $n$ requires $\delta_{n}$ contiguous time slots belonging to a scheduling
interval of $\mathcal{H}_{n}$ with a predefined pattern of operating energy
levels\footnote{When the time slot duration is an hour (i.e., $H{=}24$), the
  energy level becomes a power level.}  $\gamma^{op}_{n}(h)$ such that
$\gamma^{op}_{n}(h){>}\gamma^{min}_{n}~\forall h{\in}[0,~\delta_{n}{-}1]$ and
$\gamma^{op}_{n}(h){=}\gamma^{min}_{n}$ otherwise. Clearly,
$\beta_{n}{\geq}\alpha_{n}{+}\delta_{n}{-}1$. Without loss of generality, we set
$\gamma^{min}_{n}$ to 0; the resulting daily energy consumption $E_{n}$,
therefore, is given by
\begin{equation}
  \label{eq:daily_energy_consumption}
  E_{n} = \sum_{h=0}^{\delta_{n}-1} \gamma^{op}_{n}\left(h\right) .
\end{equation}
Fig.~\ref{fig:op_energy_level_example} shows an example of operating energy
levels.

Note that, if an appliance requires multiple, separate scheduling intervals
(e.g., one for 6 AM--11 AM and the other for 1 PM--5 PM), it can be modeled as
multiple (virtual) appliances, each of them having only one contiguous
scheduling interval (i.e., appliance 1 with 6 AM--11 AM and appliance 2 with 1
PM--5 PM).

A simple and straightforward formulation of atomic scheduling is using the
starting times of operations as optimization variables, i.e.,
\begin{equation}
  \label{eq:start_time_vector}
  \mathbf{s} \triangleq \left[s_{1},\ldots,s_{N}\right]
\end{equation}
where a feasible set of starting times for user $n$ is given by
\begin{equation}
  \label{eq:atomic_feasible_set_start_times}
  \mathcal{S}_{n} \triangleq \left\{s_{n} \big| s_{n}=i \bmod H,~\forall i{\in} [\alpha_{n},~\beta_{n}{-}\delta_{n}{+}1]\right\}
\end{equation}
The total load across all users at each time slot $h$, therefore, can be
expressed in terms of starting times
$s_{n}{\in}\mathcal{S}_{n},~\forall n{\in}\mathcal{N}$, as follows
\begin{equation}
  \label{eq:total_hourly_load1}
  L_{h}(\mathbf{s}) \triangleq \sum_{n \in \mathcal{N}}
  \gamma^{op}_{n}\left(\left(h-s_{n}\right)\bmod H\right) I_{\mathcal{R}_{n}(s_{n})}(h)
\end{equation}
where $\mathcal{R}_{n}(s_{n})$ is a range of user $n$'s appliance operation
for $s_{n}$ defined
by
\begin{equation}
  \label{eq:s_range}
  \mathcal{R}_{n}(s_{n}) \triangleq \left\{h \big| h=i \bmod H,~\forall i{\in} \left[s_{n},~s_{n}+\delta_{n}-1\right]\right\}
\end{equation}
and $I_{\mathcal{R}_{n}(s_{n})}(h)$ is an indicator function; note that, for a
set $\mathcal{A}$, the indicator function is defined as
\begin{equation}
  \label{eq:indicator_function}
  I_{\mathcal{A}}(a) \triangleq  \left\{
    \begin{array}{ll}
      1 & \mbox{if } a \in \mathcal{A}, \\
      0 & \mbox{otherwise}.
    \end{array}
  \right.
\end{equation}

Because the feasible set is now discrete, we have to evaluate the objective
function for all the elements in the feasible set, which makes the optimization
problem impractical for large $N$ and $H$. For instance, when $N{=}100$ and
$H{=}24$ with the worst case scenario of $\alpha_{n}{=}0$, $\beta_{n}{=}23$, and
$\delta_{n}{=}1$ for all $n{\in}\mathcal{N}$, global optimization by direct
enumeration would require evaluating the objective function $24^{100}$ times,
which is on the order of $10^{138}$ times.
To address the scalability issue, therefore, we provide an alternative
formulation of atomic scheduling based on the optimal routing framework, which
is amenable to convex relaxation and enables systematic analysis of suboptimal
solutions with upper and lower bounds.

\subsection{Optimal Routing-Based Formulation}
\label{sec:optimal-routing-formulation}
We provide a new formulation of the optimal scheduling of appliance energy
consumption based on the framework of \textit{optimal routing} in networking
\cite[Ch.~5]{bertsekas92:_data}; this new formulation takes into account the
atomicity of appliance operations in scheduling without taking any additional
objective functions.

As for optimization variables, instead of the energy consumption scheduling
vector $\mathbf{x}_{n}$ in \eqref{eq:energy_consumption_vector}, we use
\textit{flow configurations of users over multiple paths} between the common
source and destination nodes of the network shown in
Fig.~\ref{fig:atomic_scheduling}~(a)\footnote{This is for the case of hourly
  time slots, i.e., $H{=}24$.}, which are defined as follows:
\begin{figure}[!tb]
  \begin{center}
    \includegraphics[width=.8\linewidth]{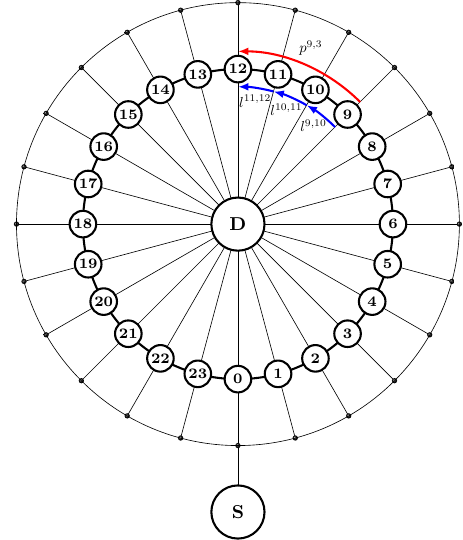}\\%
    {\scriptsize (a)}\\
    \vspace{0.2cm}
    \includegraphics[width=.8\linewidth]{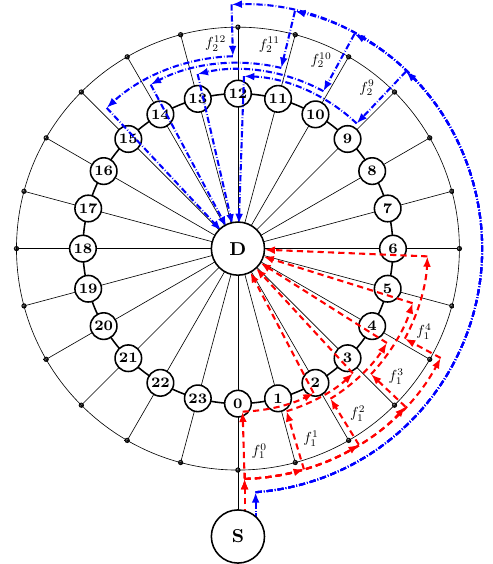}\\%
    {\scriptsize (b)}
  \end{center}
  \caption{Atomic energy consumption scheduling based on optimal routing: (a) A
    network connecting the source ($\mathbf{S}$) and the destination
    ($\mathbf{D}$) through 24 intermediate nodes with a sample path ($p^{9,3}$)
    and its constituent links ($l^{9,10}$, $l^{10,11}$, and $l^{11,12}$); (b)
    mapping of all possible atomic operations of two appliances into two groups
    of flows ($f_{1}^{0},\ldots,f_{1}^{4}$ and $f_{2}^{9},\ldots,f_{2}^{12}$)
    over multiple paths on the network.}
  \label{fig:atomic_scheduling}
\end{figure}

The users belonging to the set $\mathcal{N}$ share a set $\mathcal{P}$ of paths
connecting the source node $\mathbf{S}$ and the destination node $\mathbf{D}$,
where the path set $\mathcal{P}$ is defined as
\begin{equation}
  \label{eq:path_set}
  \mathcal{P} \triangleq \left\{p^{i,j}\Big|i=0,\ldots,H{-}1,~j=1,\ldots,H{-}1\right\}
\end{equation}
where $i$ and $j$ are the starting node number on the ring of a path and the
number of hops, respectively; for instance, as illustrated in
Fig.~\ref{fig:atomic_scheduling}~(a), $p^{9,3}$ is a three-hop path consisting
of links $l^{9,10}$, $l^{10,11}$, and $l^{11,12}$ on the ring. Note that we do
not take into account the radial links connecting the source and the destination
nodes to and from the intermediate nodes on the ring in calculating hop counts
and path/link costs.\footnote{This means that their link costs are zero.}  We
now define
$f_{n}^{s}$ as a flow that user $n$ sends on path $p^{s,\delta_{n}}$, which
represents an atomic operation of user $n$'s appliance with starting time slot
$s$ and duration of $\delta_{n}$ slots. Fig.~\ref{fig:atomic_scheduling}~(b)
shows mapping of all possible atomic appliance operations of two users (i.e.,
$\alpha_{1}{=}0$, $\beta_{1}{=}5$, $\delta_{1}{=}2$ for user 1 and
$\alpha_{2}{=}9$, $\beta_{2}{=}14$, $\delta_{2}{=}3$ for user 2) into
corresponding groups of flows (i.e., $f_{1}^{0},{\ldots},f_{1}^{4}$ for user 1
and $f_{2}^{9},{\ldots},f_{2}^{12}$ for user 2).

We use the flow configurations of all users as the optimization variables, which
are defined as follows:
\begin{equation}
  \label{eq:user_flow_configuration}
  \mathbf{f} \triangleq \left[\boldsymbol{f}_{1},\ldots,\boldsymbol{f}_{n},\ldots,\boldsymbol{f}_{N}\right]
\end{equation}
where
\begin{equation}
  \label{eq:user_flow_vector}
  \boldsymbol{f}_{n} \triangleq \left[f^{0}_{n},\ldots,f^{H-1}_{n}\right] .
\end{equation}
A feasible atomic energy consumption scheduling set for user $n$ is given by
\begin{equation}
  \label{eq:atomic_feasible_set}
  \begin{split}
    \mathcal{F}_n{=} \Bigg\{ & \mathbf{f}_n \Bigg| \sum_{s \in
      \mathcal{S}_{n}}f^{s}_{n}{=}1, \\
    & f^{s}_{n}{\in}\left\{0,~1\right\},~\forall
    s{\in}\mathcal{S}_{n},~f^{s}_{n}{=}0,~\forall
    s{\in}\mathcal{H}\backslash\mathcal{S}_{n}\Bigg\}
  \end{split}
\end{equation}
where $\mathcal{S}_{n}$ is the feasible set of starting times for user $n$ that
is already defined in \eqref{eq:atomic_feasible_set_start_times} but in terms of
$s$ instead of $s_{n}$. Note that the constraint of
$\sum_{s{\in}\mathcal{S}_{n}}f^{s}_{n}{=}1$ in \eqref{eq:atomic_feasible_set}
ensures that an optimal user flow configuration vector $\boldsymbol{f}_{n}$ has
only one non-zero element representing the starting time of a scheduled atomic
operation.

With the user flow configurations, the total load across all users at each time
slot $h{\in}\mathcal{H}$ can be calculated as the sum of all flows passing over
the link $l^{h,\left((h+1){\bmod}H\right)}$, i.e.,
\begin{equation}
  \label{eq:total_hourly_load2}
  L_{h}(\mathbf{f}) \triangleq \sum_{n \in
    \mathcal{N}}\gamma^{op}_{n} \left(\left(h{-}s\right){\bmod}H\right) \left(\sum_{s \in
      \mathcal{S}_{n}}
    f^{s}_{n}I_{R_{n}(s)}(h)\right) .
\end{equation}
Now that we have an expression of the total load at each time slot in terms of
flows representing atomic operations of appliances, we can formulate the problem
of optimal scheduling of appliance energy consumption for a given objective
function of $\phi\left(\boldsymbol{L}\left(\mathbf{f}\right)\right)$ with
$\boldsymbol{L}\left(\mathbf{f}\right) {\triangleq}
\left[L_{0}\left(\mathbf{f}\right),\ldots,L_{H-1}\left(\mathbf{f}\right)\right]$
as follows:
\begin{equation}
  \label{eq:atomic_optimal_scheduling}
  \underset{\mathbf{f}_n \in \mathcal{F}_{n},\forall n \in
    \mathcal{N}}{\mathbf{minimize}} ~~ \phi\left(\boldsymbol{L}\left(\mathbf{f}\right)\right) .
\end{equation}

For a convex objective function, this problem becomes a \textit{Boolean-convex}
problem, since the sum constraint in \eqref{eq:atomic_feasible_set} is linear
and the optimization variable $f^{s}_{n}$ is restricted to take only 0 or 1 as
in \cite{joshi09:_sensor_selec_convex_optim}. Note that this problem still
cannot be solved efficiently but, unlike the problem formulation directly based
on starting times in \eqref{eq:start_time_vector}, is amenable to the successive
convex relaxation technique that we describe in
Section~\ref{sec:successive_convex_relaxation}.

\subsection{Successive Convex Relaxation}
\label{sec:successive_convex_relaxation}
We can obtain the \textit{convex relaxation} of the atomic optimal scheduling
problem in \eqref{eq:atomic_optimal_scheduling} by replacing the nonconvex
constraints $f^{s}_{n}{\in}\left\{0,~1\right\}$ in
\eqref{eq:atomic_feasible_set} with the convex constraints
$0{\leq}f^{s}_{n}{\leq}1$ as follows:
\begin{equation}
  \label{eq:convex_relaxation_atomic_optimal_scheduling}
  \underset{\mathbf{f}_n \in \mathcal{\hat{F}}_{n},\forall n \in
    \mathcal{N}}{\mathbf{minimize}} ~~ \phi\left(\boldsymbol{L}\left(\mathbf{f}\right)\right)
\end{equation}
where 
\begin{equation}
  \label{eq:convex_relaxation_atomic_feasible_set}
  \begin{split}
    \mathcal{\hat{F}}_{n} = \Bigg\{ & \mathbf{f}_n \Bigg| \sum_{s \in
      \mathcal{S}_{n}}f^{s}_{n}{=}1, \\
    & 0 \leq f^{s}_{n} \leq 1,~\forall
    s{\in}\mathcal{S}_{n},~f^{s}_{n}{=}0,~\forall
    s{\in}\mathcal{H}\backslash\mathcal{S}_{n}\Bigg\} .
  \end{split}
\end{equation}

For a convex objective function, this problem becomes convex because the new
feasibility set $\mathcal{\hat{F}}_{n}$ is now convex where all the equality and
inequality constraints on $\mathbf{f}$ are linear. This convex-relaxed problem
in \eqref{eq:convex_relaxation_atomic_optimal_scheduling}, therefore, can be
solved efficiently, for instance, using the well-known interior-point method
\cite{boyd04:_convex}.

Let $\mathbf{\hat{f}}$ denote a solution to the relaxed problem in
\eqref{eq:convex_relaxation_atomic_optimal_scheduling}. The relaxed atomic
scheduling problem in \eqref{eq:convex_relaxation_atomic_optimal_scheduling} is
not equivalent to the original problem in \eqref{eq:atomic_optimal_scheduling}
because the elements of the optimal solution $\mathbf{\hat{f}}$ can take
fractional values (e.g., 0.75). The optimal objective value of the relaxed
atomic scheduling problem in
\eqref{eq:convex_relaxation_atomic_optimal_scheduling}, however, provides a
lower bound on the optimal objective value of the original problem in
\eqref{eq:atomic_optimal_scheduling}; the optimal value of the relaxed problem
is less than or equal to that of the original problem because the feasible set
for the relaxed problem contains the feasible set for the original problem.

We can also use the solution of the relaxed problem in
\eqref{eq:convex_relaxation_atomic_optimal_scheduling} to generate a suboptimal
solution. Note that we cannot simply choose $N$ largest elements of the
optimization vector as is done for the optimal selection of sensor measurements
in \cite{joshi09:_sensor_selec_convex_optim}; because the optimization variable
$\mathbf{f}$ is a vector of vectors (i.e., $\boldsymbol{f}_{n}$), the selection
of largest elements should be done per component vector. Also, we found that the
convex relaxation spreads the starting times of appliance energy consumption
over the same scheduling interval when there are several appliances with
identical parameter values (i.e., $\alpha_{n}$, $\beta_{n}$,
$\gamma_{n}^{op}(\cdot)$, and $\delta_{n}$), which often results in many
elements with the same values in the component vectors. If we choose randomly
one element among the same-valued ones to break ties in this case, the
performance of resulting suboptimal solution is not close to the optimal one.

Therefore, here we present a new technique, called \textit{successive convex
  relaxation}, where we iteratively drop fractional-valued elements and apply
convex relaxation to the resulting problem until we find a feasible
solution. This technique is similar to the cutting-plane algorithm in mixed
integer linear programming \cite[Ch.~9]{bradley77:_applied} in that, at each
iterative step, new constraints are added to refine the feasible region. The
proposed technique, however, is much simpler in forming new constraints where it
does not introduce any new slack variables and takes into account the structure
of the optimization variables. The detailed procedure is described in
Algorithm~\ref{alg:successive_convex_relaxation}, where we introduce two
variables, i.e., $\theta_{D}$, the threshold value for dropping, and $N_{D}$,
the maximum number of fractional-valued elements that can be dropped per
iteration. Note that the overall complexity of the interior-point method to
solve the relaxed convex optimization problem is
$\mathcal{O}\left(\left(NH\right)^{3}\right)$ operations \cite{boyd04:_convex},
which is needed per iteration and a dominating factor of the complexity of the
proposed successive convex relaxation technique. With these variables,
therefore, we can do a fine control of the number of fractional-valued elements
dropped per iteration and the number of iterations to finish the procedure: A
reasonable value of $\theta_{D}$ (e.g., 0.1) can prevent many high
fractional-valued elements (e.g., ${\geq}0.5$) from being dropped unnecessarily
per iteration when $N_{D}$ is set to a rather large number. On the other hand,
when a relaxed convex optimization problem gives a solution with a large number
of fractional-valued elements smaller than $\theta_{D}$, we can drop them up to
$N_{D}$. In this way we can speed up the whole procedure while not sacrificing
the quality of the resulting suboptimal solution.
\SetStartEndCondition{ }{}{}%
\SetKwProg{Fn}{}{\{}{}\SetKwFunction{FRecurs}{void FnRecursive}%
\SetKwFor{For}{for}{}{}%
\SetKwIF{If}{ElseIf}{Else}{if}{}{elif}{else}{}%
\SetKwFor{While}{while}{}{}%
\SetKwRepeat{Repeat}{repeat\{}{until}%
\AlgoDontDisplayBlockMarkers\SetAlgoNoEnd\SetAlgoNoLine%
\DontPrintSemicolon%
\SetKwFunction{BL}{ByteLength}%
\SetKwFunction{CF}{Conform}%
\SetKwFunction{DR}{Drop}%
\SetKwFunction{EA}{Estimate$\alpha$}%
\SetKwFunction{EB}{ExcessBW}%
\SetKwFunction{EQ}{Enque}%
\SetKwFunction{ER}{EstimateRate}%
\SetKwFunction{LN}{Length}%
\SetKwFunction{MN}{Min}%
\SetKwFunction{MX}{Max}%
\SetKwFunction{UR}{UniformRandom}%
\IncMargin{0.3em}%
\LinesNumbered%
\SetCommentSty{emph}%
\SetAlFnt{\small}%
\begin{algorithm}[!t]
  $\mathcal{D} \leftarrow \emptyset$\tcc*[r]{initialize the set of elements to
    drop}%
  \While{\textit{true}}{%
    $\mathbf{\hat{f}} \leftarrow $Solution of
    \eqref{eq:convex_relaxation_atomic_optimal_scheduling} with the equality
    constraints replaced by
    ``$f^{s}_{n}=0,~\forall s \in
    \mathcal{H}\backslash\mathcal{S}_{n}~{\cup}~\mathcal{D}$''
    in \eqref{eq:convex_relaxation_atomic_feasible_set}\;%
    Exclude elements that are the maximum of
    $\boldsymbol{\hat{f}}_{n},~\forall n \in \mathcal{N}$, arrange in ascending
    order the remaining elements of $\mathbf{\hat{f}}$ whose values are less
    than one, and let $i_{1},i_{2},\ldots$ denote their indexes\;%
    $\mathbf{\hat{f}}_{i_{1}} \leftarrow 0$\tcc*[r]{always drop the
      minimum-valued element}%
    $\mathcal{D} \leftarrow \mathcal{D} \cup {i_{1}}$\;%
    \For{$j=2,\ldots,N_{D}$}{%
      \eIf{$\mathbf{\hat{f}}_{i_{j}}<\theta_{D}$}{%
        $\mathbf{\hat{f}}_{i_{j}} \leftarrow 0$\tcc*[r]{drop it}%
        $\mathcal{D} \leftarrow \mathcal{D} \cup {i_{j}}$\;%
      }{%
        Break\tcc*[r]{return to the while loop}%
      }%
    }%
    \If{there is only one nonzero element in
      $\boldsymbol{\hat{f}}_{n}, \forall n{\in}\mathcal{N}$}{%
      Break\tcc*[r]{stop here; solution found}%
    }%
  }
  \caption{Successive convex relaxation.}
  \label{alg:successive_convex_relaxation}
\end{algorithm}

\subsection{Examples}
\label{sec:examples}
Below we provide specific examples of the optimal atomic scheduling for
frequently used objective functions for DSM in residential smart grid.

\subsubsection{Energy Cost Minimization}
\label{sec:energy-cost-minim}
For energy cost minimization, we can formulate it as the following optimization
problem similar to \cite{mohsenian-rad10:_auton}:
\begin{equation}
  \label{eq:energy_cost_minimization}
  \underset{\mathbf{f}_n \in \mathcal{F}_{n},\forall n \in
    \mathcal{N}}{\mathbf{minimize}} ~~ \sum_{h \in \mathcal{H}} C_{h}\left(
    L_{h}(\mathbf{f})
  \right)
\end{equation}
where $C_{h}(\cdot)$ is a cost function indicating the cost of generating or
distributing electricity energy by the energy source at a time slot $h$.

Replacing the feasible set $\mathcal{F}_{n}$ with $\mathcal{\hat{F}}_{n}$
in \eqref{eq:energy_cost_minimization}, we obtain the relaxed energy cost
minimization problem as follows:
\begin{equation}
  \label{eq:convex_relaxation_energy_cost_minimization}
  \underset{\mathbf{f}_n \in \mathcal{\hat{F}}_{n},\forall n \in
    \mathcal{N}}{\mathbf{minimize}} ~~ \sum_{h \in \mathcal{H}} C_{h}\left(
    L_{h}(\mathbf{f})
  \right) .
\end{equation}

\subsubsection{Peak-to-Average Ratio Minimization}
\label{sec:peak-average-ratio}
Considering the total energy consumption $\sum_{n \in \mathcal{N}}E_{n}$ is
fixed, we can formulate the PAR minimization problem as follows:
\begin{equation}
  \label{eq:par_minimization1}
  \underset{\mathbf{f}_n \in \mathcal{F}_{n},~\forall n \in
    \mathcal{N}}{\mathbf{minimize}} ~~ \max_{h \in \mathcal{H}} \left(
    L_{h}(\mathbf{f})
  \right) .
\end{equation}
Note that \eqref{eq:par_minimization1} is difficult to directly solve due to the
$\max(\cdot)$ term in the objective function. As mentioned in
\cite{mohsenian-rad10:_auton}, however, this optimization problem can be turned
into a Boolean-linear program, a special case of Boolean-convex optimization, by
introducing a new auxiliary variable $\Gamma$ as follows:
\begin{equation}
  \label{eq:par_minimization2}
  \begin{split}
    \underset{\Gamma,\mathbf{f}_n{\in}\mathcal{F}_{n},{\forall}n{\in}
      \mathcal{N}}{\mathbf{minimize}} & ~~ \Gamma \\
    \mathbf{subject~to~} & ~~ \Gamma \geq L_{h}(\mathbf{f}),~
    \forall h \in \mathcal{H} .
  \end{split}
\end{equation}

Again, replacing the feasible set $\mathcal{F}_{n}$ with $\mathcal{\hat{F}}_{n}$
in \eqref{eq:par_minimization2}, we obtain the relaxed PAR minimization problem,
i.e.,
\begin{equation}
  \label{eq:convex_relaxation_par_minimization2}
  \begin{split}
    \underset{\Gamma,\mathbf{f}_n{\in}\mathcal{\hat{F}}_{n},{\forall}n{\in}
      \mathcal{N}}{\mathbf{minimize}} & ~~ \Gamma \\
    \mathbf{subject~to~} & ~~ \Gamma \geq L_{h}(\mathbf{f}),~
    \forall h \in \mathcal{H} .
  \end{split}
\end{equation}
Like the convex optimization, the linear program can be efficiently solved by
either the simplex method or the interior-point method
\cite{boyd04:_convex}. Note that, when applying the successive convex relaxation
technique described in Algorithm~\ref{alg:successive_convex_relaxation}, the
auxiliary variable $\Gamma$ is not subject to dropping unlike $\mathbf{f}$.


\section{Numerical Results}
\label{sec:numerical_results}
We demonstrate the application of the successive convex relaxation technique to
the atomic scheduling of appliance energy consumption through numerical examples
for the minimization of total energy cost and PAR in the aggregated load. In all
the examples, a day is divided into 24 time slots (i.e., $H{=}24$), and each
user is to have an appliance randomly selected from the appliances whose energy
consumption requirements are summarized in Table~\ref{tbl:appliance_parameters};
for simplicity, we assume constant operating energy levels for all appliances.
\begin{table}
  \begin{threeparttable}
    \centering
    \caption{Appliance energy consumption requirements (adapted from
      \cite{mohsenian-rad10:_auton})}
    \label{tbl:appliance_parameters}
    \begin{tabular}{|c|r|r|r|r|}
      \hline
      \multirow{2}{*}{Appliance} & \multicolumn{4}{|c|}{Parameters} \\ \cline{2-5}
      & $\alpha$ [h] & $\beta$ [h] & $\gamma^{op}$ [kWh] & $\delta$ [h] \\ \hline\hline
      Dish Washer & 0 & 23 & 0.7200 & 2 \\ \hline
      Washing Machine (Energy-Star) & 0 & 23 & 0.4967 & 3 \\ \hline
      Washing Machine (Regular) & 0 & 23 & 0.6467 & 3 \\ \hline
      Clothes Dryer & 0 & 23 & 0.6250 & 4 \\ \hline
      Plug-in Hybrid Electric Vehicle & \multirow{2}{*}{22\tnote{*}} &
      \multirow{2}{*}{29\tnote{*}} & \multirow{2}{*}{3.3000} & \multirow{2}{*}{3} \\
      (PHEV) & & & & \\ \hline
    \end{tabular}
    \begin{tablenotes}
    \item[1] Scheduling interval of 10 PM--5 AM.
    \end{tablenotes}
  \end{threeparttable}
\end{table}

For the energy cost minimization problem described in
Section~\ref{sec:energy-cost-minim}, we assume a simple quadratic hourly cost
function as in \cite{mohsenian-rad10:_auton}, which is given by
\begin{equation}
  \label{eq:hourly_cost_function}
  C_{h}\left(L_{h}\right) = a_{h}L_{h}^{2} ~ \mbox{[cent]}
\end{equation}
where
\begin{equation}
  \label{eq:hourly_cost_function_coefficient}
  a_{h} = \left\{
    \begin{array}{ll}
      0.2 & \mbox{if $h \in [0,~7]$},\\
      0.3 & \mbox{if $h \in [8,~23]$}.
    \end{array}
  \right.
\end{equation}
The objective function in \eqref{eq:convex_relaxation_energy_cost_minimization}
for this hourly cost function can be expressed as
\begin{equation}
  \label{eq:convex_relaxation_energy_cost_function}
  \psi\left(\mathbf{f}\right) =
  \phi\left(\boldsymbol{L}\left(\mathbf{f}\right)\right)
  = \sum_{h \in \mathcal{H}} a_{h}\left(L_{h}\left(\mathbf{f}\right)\right)^2 .
\end{equation}
Then its first and second derivatives, which are needed for convex optimization,
are given as follows: For $n{\in}\mathcal{N}$ and $s{\in}\mathcal{H}$, the first
derivative (i.e., the \textit{gradient} of $\psi(\mathbf{f})$) is given by
\begin{equation}
  \label{eq:convex_relaxation_energy_cost_gradient}
  \begin{split}
    \frac{\partial \psi(\mathbf{f})}{\partial f^{s}_{n}} & = \\
    & 2 \sum_{h \in \mathcal{H}} a_{h} \gamma^{op}_{n}
    \left(\left(h{-}s\right){\bmod}H\right)
    L_{h}\left(\mathbf{f}\right)I_{\mathcal{R}_{n}(s)}(h) .
  \end{split}
\end{equation}
For $n_{1},n_{2}{\in}\mathcal{N}$ and $s_{1},s_{2}{\in}\mathcal{H}$, the second
derivative (i.e., the \textit{Hessian} of $\psi(\mathbf{f})$), which is positive
due to the convexity of the objective function in
\eqref{eq:convex_relaxation_energy_cost_function}, is given by
\begin{equation}
  \label{eq:convex_relaxation_energy_cost_hessian}
  \begin{split}
    & \frac{\partial^{2} \psi(\mathbf{f})}{\partial f^{s_{1}}_{n_{1}} \partial
      f^{s_{2}}_{n_{2}}} = \\
    & 2 \sum_{h \in \mathcal{H}} a_{h}
    \gamma^{op}_{n_{1}}\left(\left(h{-}s_{1}\right){\bmod}H\right)
    \gamma^{op}_{n_{2}}\left(\left(h{-}s_{2}\right){\bmod}H\right) \\
    & \times I_{\mathcal{R}_{n_{1}}(s_{1})}(h)
    I_{\mathcal{R}_{n_{2}}(s_{2})}(h).
  \end{split}
\end{equation}

\begin{figure}[!tb]
  \begin{center}
    \includegraphics[width=.8\linewidth]{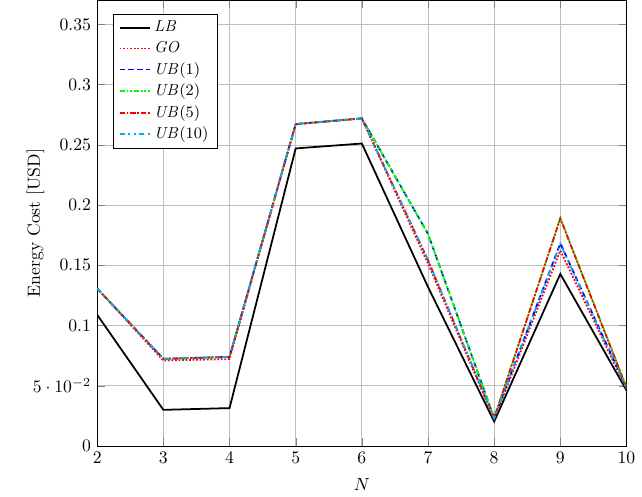}\\%
    {\scriptsize (a)}\\
    \includegraphics[width=.8\linewidth]{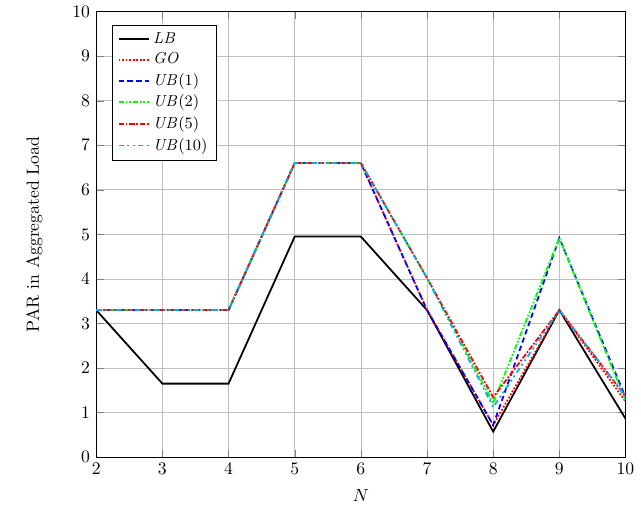}\\%
    {\scriptsize (b)}
  \end{center}
  \caption{Comparison of upper (UB) and lower bounds (LB) with true values from
    global optimization (GO) in atomic energy consumption scheduling for (a)
    cost minimization and (b) PAR minimization.}
  \label{fig:bounds_comparison}
\end{figure}
To evaluate the performance of the successive convex relaxation technique, we
first obtain the lower bound ($\mathit{LB}$) using
\eqref{eq:convex_relaxation_energy_cost_minimization} and
\eqref{eq:convex_relaxation_par_minimization2} for the minimization of energy
cost and PAR, respectively. Then we obtain suboptimal solutions, which are also
upper bounds, using Algorithm~\ref{alg:successive_convex_relaxation} with the
dropping threshold $\theta_{D}$ fixed to 0.1 and different values of $N_{D}$
($\mathit{UB}(N_{D})$). We compare the lower and the upper bounds with the
optimal objective value from global optimization ($\mathit{GO}$) using direct
enumeration. The results are shown in Fig.~\ref{fig:bounds_comparison}, where,
due to the huge size of the feasible set for direct enumeration given in
\eqref{eq:atomic_feasible_set}, we limit the maximum value of $N$ to 10; for
$N{=}10$, the size of feasible set is about $2.3{\times}10^{13}$, and in case of
the energy cost minimization, it took 105.2 hours (i.e., more than 4 days) to
obtain the optimal solution from direct enumeration using an OpenMP-based
parallelized version of C++ program on a workstation with two
Intel\textsuperscript{\textregistered} Xeon\textsuperscript{\textregistered}
processors running at 2.3 GHz providing 20 cores and 40 threads in total, while
it took 23.3 seconds to obtain the suboptimal solution from the successive
convex relaxation with $N_{D}{=}1$ (requiring 207 iterative steps of convex
relaxation) using a MATLAB\textsuperscript{\textregistered} script with OPTI
toolbox \cite{currie12:_opti} on the same machine. Note that due to the random
selection of appliances and their different requirements for energy consumption,
the resulting energy cost and PAR are not proportional to $N$.

The results in Fig.~\ref{fig:bounds_comparison} show that both lower and upper
bounds are very close to the true optimal values. In case of the upper bounds,
they are even identical to the true optimal values. For instance, the upper
bounds on the energy cost for all values of $N_{D}$ when $N{=}2$ and for
$N_{D}{=}1,2,5$ when $N{=}5$ are identical to true optimal values; in case of
PAR, the upper bounds for all values of $N_{D}$ when $N{\leq}6$ are identical to
true optimal values. Considering the huge difference in the computational
complexity between the two approaches, these results are remarkable. Of the
results for energy cost in Fig.~\ref{fig:bounds_comparison}~(a) and PAR in
Fig.~\ref{fig:bounds_comparison}~(b), we found that the integrality gap in
convex relaxation \cite{chlamtac12:_convex_relax_integ_gaps} is more visible for
the lower bounds on the PAR: Note that, unlike the total cost whose calculation
involves all the appliances, the PAR depends on less number of appliances in its
calculation, i.e., only those that contribute to a specific time slot whose load
is maximum. The relaxation process, however, makes more appliances contribute to
the PAR calculation by spreading fractional-valued flows over all possible
paths.

\begin{figure*}[!htb]
  \begin{minipage}{.325\linewidth}
    \begin{center}
      \includegraphics[width=\linewidth]{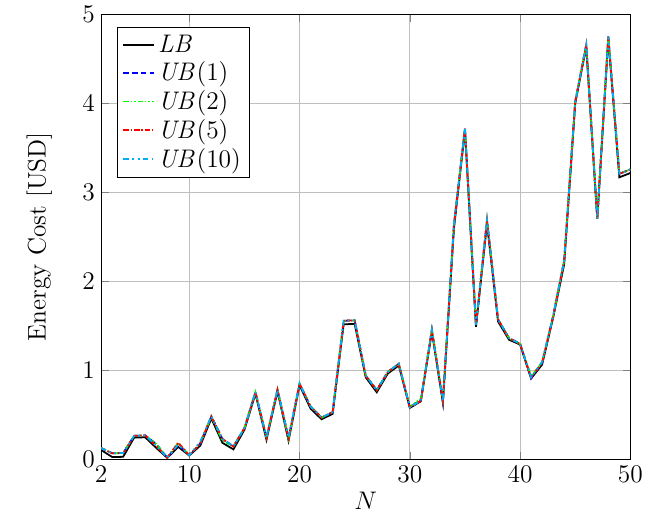}\\%
      {\scriptsize (a)}
    \end{center}
  \end{minipage}
  \hfill
  \begin{minipage}{.325\linewidth}
    \begin{center}
      \includegraphics[width=\linewidth]{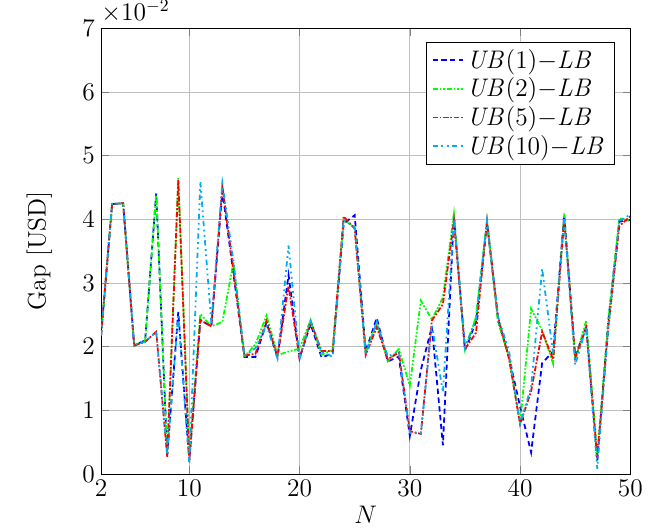}\\%
      {\scriptsize (b)}
    \end{center}
  \end{minipage}
  \hfill
  \begin{minipage}{.325\linewidth}
    \begin{center}
      \includegraphics[width=\linewidth]{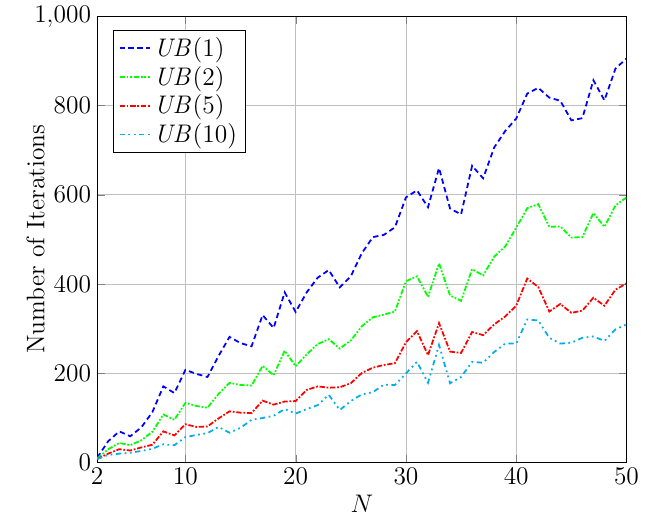}\\%
      {\scriptsize (c)}
    \end{center}
  \end{minipage}
  \caption{Atomic energy consumption scheduling for cost minimization:
    (a) Upper ($\mathit{UB}$) and lower bounds ($\mathit{LB}$); (b) gaps; (c)
    number of iterations.}
  \label{fig:cost_min_successive_convex_relaxation}
\end{figure*}
\begin{figure*}[!htb]
  \begin{minipage}{.325\linewidth}
    \begin{center}
      \includegraphics[width=\linewidth]{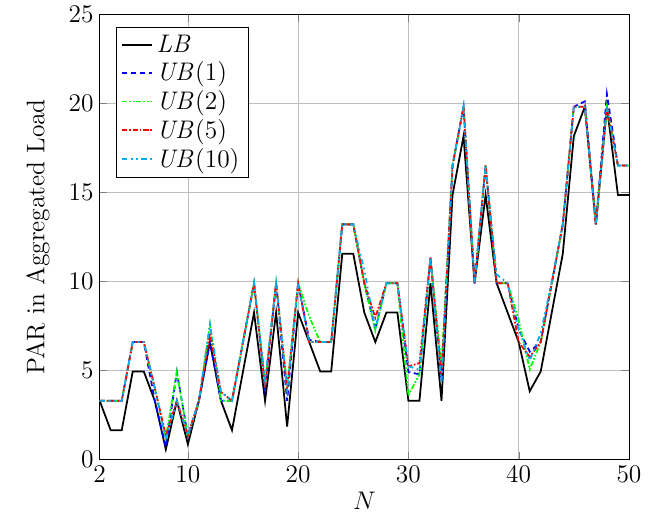}\\%
      {\scriptsize (a)}
    \end{center}
  \end{minipage}
  \hfill
  \begin{minipage}{.325\linewidth}
    \begin{center}
      \includegraphics[width=\linewidth]{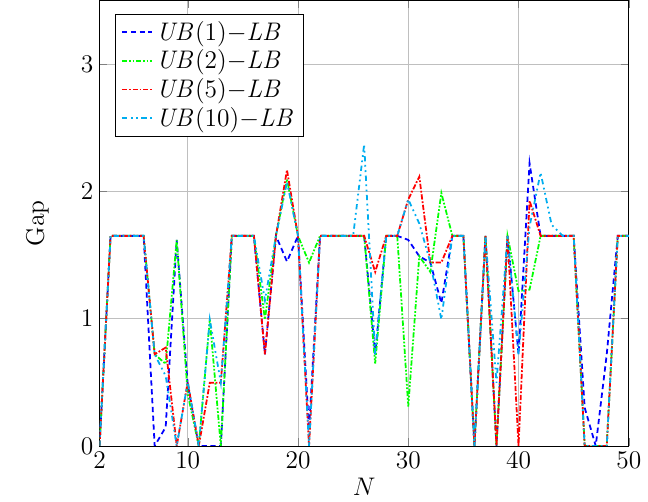}\\%
      {\scriptsize (b)}
    \end{center}
  \end{minipage}
  \hfill
  \begin{minipage}{.325\linewidth}
    \begin{center}
      \includegraphics[width=\linewidth]{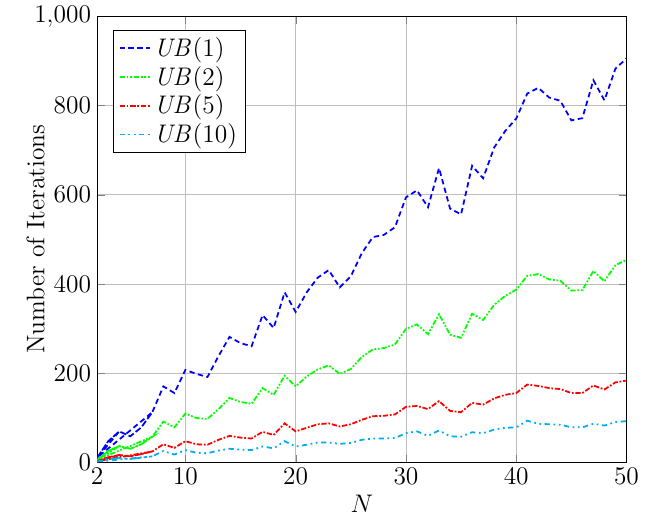}\\%
      {\scriptsize (c)}
    \end{center}
  \end{minipage}
  \caption{Atomic energy consumption scheduling for PAR minimization:
    (a) Upper ($\mathit{UB}$) and lower bounds ($\mathit{LB}$); (b) gaps; (c)
    number of iterations.}
  \label{fig:par_min_successive_lp_relaxation}
\end{figure*}

To further investigate the quality of lower and upper bounds and the impact of
different values of $N_{D}$ on the performance of successive convex relaxation
technique, we obtain the lower and the upper bounds, their gaps defined as the
difference between upper and lower bounds, and the number of iterations for
upper bounds for the value of $N$ from 2 to
50. Figs.~\ref{fig:cost_min_successive_convex_relaxation} and
\ref{fig:par_min_successive_lp_relaxation} show the results for energy cost
minimization and PAR minimization, respectively.

From the figures, we observe that the overall trend in the results from the
minimization of energy cost and PAR are similar to each other, except the
relatively larger gaps in the PAR minimization that we already discussed with
the results shown in Fig.~\ref{fig:bounds_comparison}. For both minimization
problems, the upper bounds are quite close to the lower bound for a broader
range of values of $N$, and the curves for lower bounds with different values of
$N_{D}$ are hardly distinguishable. The gaps shown in
Fig.~\ref{fig:cost_min_successive_convex_relaxation}~(b) and
Fig.~\ref{fig:par_min_successive_lp_relaxation}~(b) further confirm that the
quality of suboptimal solutions represented by the lower bounds hardly depends
on the values of $N_{D}$ but slightly improves as $N$ increases; even though
both upper and lower bound values roughly increase as $N$ increases, the gaps
fluctuate but do not show any trend of increasing.

While Fig.~\ref{fig:cost_min_successive_convex_relaxation} (a) and (b) and
Fig.~\ref{fig:par_min_successive_lp_relaxation} (a) and (b) show that $N_{D}$
does not have any visible impact on the quality of suboptimal solutions,
Fig.~\ref{fig:cost_min_successive_convex_relaxation}~(c) and
Fig.~\ref{fig:par_min_successive_lp_relaxation}~(c) show that increasing the
value of $N_{D}$ can significantly reduce the number of iterations. As discussed
in Section~\ref{sec:successive_convex_relaxation}, these results show that a
larger value of $N_{D}$, in combination with a lower dropping threshold value
(i.e., $\theta_{D}=0.1$), can improve the speed of the successive convex
relaxation while maintaining the quality of resulting suboptimal solutions.

\section{Conclusions}
\label{sec:conclusions}
\balance                        
The atomicity of appliance operation has never been given serious attention in
scheduling of appliance energy consumption in autonomous DSM for residential
smart grid. The current dominant approach based on the vector of appliance's
hourly energy consumption may result in several gaps in scheduled appliance
energy consumption and have problems in providing enough power for appliances to
carry out required operations, which, in most cases, are atomic.

In this paper, therefore, we have provided a new formulation of appliance energy
consumption scheduling based on the optimal routing framework, which guarantees
the atomicity of resulting scheduled energy consumption as such without
additional objective functions. Compared to the straightforward problem
formulation based on the vector of possible starting times of an appliance
operation, the optimal-routing-based formulation provides a Boolean-convex
problem for a convex objective function, which is amenable to the successive
convex relaxation technique where we can apply well-known interior-point methods
for the efficient solution of relaxed convex optimization problem. Unlike
approaches based on heuristics like genetic algorithms \cite{Goldberg:89},
convex relaxation enables us to carry out systematic analysis of the original
problem with both upper and lower bounds.

The numerical results for the cost and PAR minimization problems demonstrate
that the proposed successive convex relaxation technique can provide tight upper
and lower bounds and, therefore, suboptimal solutions very close to optimal
solutions, often identical to true optimal values, in an efficient way. The
results also show that, using two control parameters, i.e., $N_{D}$ for the
maximum number of fractional-valued elements that can be dropped per iteration
and $\theta_{D}$ for a dropping threshold, we can strike the right balance
between the quality of suboptimal solutions and the number of iterations to
obtain them in applying the successive convex relaxation technique. Considering
that the original problem of atomic scheduling is a very difficult combinatorial
problem to solve using direct enumeration due to its huge size of the feasible
set, the proposed successive convex relaxation technique makes it practical to
implement the atomic scheduling of appliance energy consumption for autonomous
DSM in residential smart grid.

Note that our major focus in this paper is on the formulation of atomic
scheduling problem and efficient solution techniques based on convex relaxation
with adaptive dropping of fractional-valued elements. The extension to
distributed atomic energy consumption scheduling and advanced techniques
refining the feasible region at each iterative step to reduce the total number
of iterations are interesting topics for further study.

\section*{Acknowledgment}
This work was supported by the Centre for Smart Grid and Information Convergence
(CeSGIC) and the Research Development Fund (RDF) of Xi'an Jiaotong-Liverpool
University under grant reference number RDF-14-01-25.


\end{document}